\documentstyle[graphics,epsfig,floats,aps]{revtex}

\begin{document}
\draft \catcode`\@=11 \catcode`\@=12
\twocolumn[\hsize\textwidth\columnwidth\hsize\csname@twocolumnfalse\endcsname
\title {Exploring complex networks by walking on them}
\author{Shi-Jie Yang}
\address{Department of Physics, Beijing Normal University,
Beijing 100875, China}
\date{\today}
\maketitle

\begin{abstract}
We carry out a comparative study on the problem for a walker searching on several typical complex networks. The
search efficiency is evaluated for various strategies. Having no knowledge of the global properties of the
underlying networks and the optimal path between any two given nodes, it is found that the best search strategy is
the self-avoid random walk. The preferentially self-avoid random walk does not help in improving the search
efficiency further. In return, topological information of the underlying networks may be drawn by comparing the
results of the different search strategies.

\end{abstract}
 ]

%\newpage \pagenumbering {arabic}

\section{introduction}
In the past few years, much scientific interest has been devoted to the characterization and modelling of a wide
range of complex systems that can be described as networks\cite{AB,Dorogovtsev}. Systems such as the
Internet\cite{Faloutsos,Caldarelli,Pastor} or the World Wide Web\cite{Albert}, social communities\cite{Strogatz},
food webs\cite{Montoya}, and biological interacting networks\cite{Wagner,Jeong,Sole} are represented as a graph,
in which nodes represent the population individuals and links the physical interactions among them. Among these
networks, most have complex topological properties and dynamical features that cannot be accounted for by
classical graph modelling. It has been demonstrated that many of these real-world networks pose small-world and
clustering property\cite{Watts}. On the other hand, scale-free (SF) degree distributions seem to emerge frequently
as dominant features governing the topology of real-world networks\cite{BA}. These global properties imply a large
connectivity heterogeneity and a short average distance between nodes, which have considerable impact on the
behavior of physical processes taking place on top of the network. A number of models have been developed to
understand the structure and functions of underlying real-world networks. For instance, scale-free networks have
been shown to be resilient to random damage\cite{AJB,Callaway,Cohen} while be fragile under intentional attacks
targeting on the nodes with high degree. It is also prone to epidemic spreading (null epidemic
threshold)\cite{PV,May,Moreno}.

While a number of recent works have concentrated on the properties
of the power-law networks and how they are dynamically generated,
another interesting problem is to find efficient algorithms for
searching within these particular kinds of
graphs\cite{Adamic,Noh}. In the most general distributed search
context, one may have very little information about the location
of the target. An interesting example is provided by the recent
emergence of peer-to-peer networks, which have gained enormous
popularity with users wanting to share their computer files. In
such networks, the name of the target file may be known, but due
to the network's {\it ad hoc} nature, the node holding the file is
not known until a real-time search is performed. File-sharing
systems that do not have a central server include GNUTELLA and
FREENET. Files are found by forwarding queries to one's neighbors
until the target is found. Recent measurements of GNUTELLA
networks\cite{Clip2} and simulated FREENET networks\cite{Hong}
show that they have power-law degree distributions.

Search processes would be optimal if one follows the shortest path
between two nodes under considerations. Among all paths connecting
two nodes, the shortest path is given by the one with the smallest
number of links. In realistic, however, the searcher does not
presumedly know the shortest path to reach the target. He even has
no idea of the general topology of the network that he is going to
search. It is important to design appropriate search strategies in
order to acquire high efficiency and, in the meantime, get an
overall idea of the underlying network. Random walk becomes
important in the extreme opposite case where only local
connectivity is known at each node. It is theoretically
interesting to probe how the structural heterogeneity affects the
nature of the diffusive and relaxation dynamics of the random
walk\cite{Adamic,Jespersen,Guimera}. In return, random walk is
also suggested to be a useful tool in studying the structure of
networks.

Much is known about random walks on both regular and random networks\cite{Barber,Hughes}. Recently, there have
been several studies of random walks on SWN's\cite{Lahtinen,Pandit} as well as on SF's\cite{Adamic,Noh}. In this
work, we systemically carry out comparative studies of random walks for several typical complex networks. We
suppose at every step, the walker only know neighbors of its present node. So if the target is at one neighbor of
the present node where the walker stays, this round of search is over. The search strategies adopted by the walker
include: random walk (RW), no-back (NB) walk, no-triangle-loop (NTL) walk, no-quadrangle-loop (NQL) walk, and
self-avoid (SA) random walk . For the RW, the walker may hop to a neighbor node by randomly taking one of the
links with equal probability. It forgets all information about its past. The NB walk implies that a random walker,
if possible, will not return to the node it situated at the previous step. Similarly, NTL and NQL random walks
mean that the walker will try to avoid walking in loops, with 3 or 4 edges, respectively, unless there is no other
choice. Finally, the SA random walk implies that the walker is more smart. It tries to avoid revisiting the node
that it has ever visited in a run of search.

Let the walker starts from one of the nodes and set in turn the other nodes as the target. For every pair of given
nodes, we perform 200 runs of simulations and take average of the search time. It is found that the search time is
only slightly dependent on the starting nodes, regardless of the large variance in node degree for some networks.
The overall mean search time is again average over the whole networks. It is found that search efficiency of each
walk strategy varies widely with the topology of the underlying networks. In general, the self-avoid random walk
is the most efficient search strategy.

We also perform a preferentially self-avoid (PSA) random walk on these networks, in which the walker is prone to a
near neighbor with a higher degree. In this case, the walker must know the degree of its present node, as well as
the degrees of its near neighbors. Contrary to one's intuition, it does not promote the efficiency of the search
processes. In many cases, it lowers the efficiency and only greatly increases the computer running time.

The paper is organized as follows. In Sec. II, we study the search processes on random graph networks. Section III
concerns SWN's and in Sec. IV, properties of scale-free networks is investigated. Summary is included in Sec. V.

\section{random-graph networks}
We define a random graph as $N$ labelled nodes and every pair of
the nodes being connected with probability $p$. Consequently the
total number of edges is a random variable with the expected value
$pN(N-1)/2$. The degree $k_i$ of a node $i$ follows a binomial
distribution with parameters $N-1$ and $p$:
\begin{equation}
P(k_i=k)=C^k_{N-1}p^k(1-p)^{N-1-k}.
\end{equation}

We perform random walks only on the largest cluster. The walker starts in turn from a node to reach a given node
on the network. Average is taken over 200 runs. We note that the search time is only slightly dependent on the
starting nodes. Figure 1 shows a log-log plot of average search times for various $p$, given a total number of
nodes $N=1000$. It exhibits a power-law relation $t\propto p^{-\gamma}$, with exponent $\gamma=1$. In fact, we
have performed all the above mention search strategies, and found the difference is rather small. It means that
for the random graph network, clustering effect and short path effect are not obvious. As a whole, a random walk
is also the optimal walk.

\section{the Watts-Strogatz model}
The small-world networks proposed by Watts-Strogatz are structures of much recent interest. It combines aspects
from regular and completely random lattices. Such structure may devised by adding, in a random way, links to an
ordered lattice. A major feature is that even at a very low density of additional links, the chemical distance
(minimal distance between two points) drastically decreases from its original value on the underlying regular
lattice. The WS model possesses inherent loops and therefore displays other properties than disordered treelike
structure, namely, it has a large clustering coefficient.

We start from $N$ sites on a ring, where each of the sites is connected by links to its four nearest and
next-nearest neighbors. For all pairs of sites ($i,k$), we add a link with probability $p$. As in Sec. II, we
perform random walks for various strategies on the WSN for $N=300$ and found great difference against the random
graph network. Figure 2 shows that the random walk is now a bad search strategy. The no-back walk,
no-triangle-loop and no-quadrangle-loop walks, as well as self-avoid walk run more efficiently. The NTL and NQL
walks eliminate repeating visits in loops for clustered nodes. Hence improvement of search efficiency for NTL and
NQL walks implies high clusters of nodes are popular in SWN's. The SA random walk can further largely decrease the
search time, indicating a small-world property of the underlying network takes important effect.

\section{scale-free networks}
A large number of real networks, including metabolic networks, the protein interaction network, the World Wide
Web, and even some social networks, exhibit the scale-free topology, in which the vertex has a power-law degree
distribution $p(k)\propto k^{-\gamma}$ typically with scaling exponent $2<\gamma <3$\cite{AB,Dorogovtsev}. With
the pioneer work of Barab\'{a}si and Albert (BA)\cite{BA}, dozens of scale-free models have been
constructed\cite{Bianconi,Krapivski,DGM,Ravasz,Holme,Klemm}. The fundamental ingredient in the BA model and its
variants are the network growth and the preferential attachment: vertices are added one after another to the
network, and edges are more prone to be connected to vertices with large $k$. Other models with non-growing
algorithm also exist\cite{Cald,Kim}. Below we will focus on two typical SF networks: one is the original BA model;
the other is a clustered SF model. We study the effects of clustering on the dynamical processes.

\subsection{The Barab\'{a}si-Albert model}
The algorithm of the BA model is following:
\begin{enumerate}
\item Growth: Starting with a small number $m_0$ of nodes, at every step, we add a new node with $m=2$ edges that
link the new node to $m$ different modes already present in the
system.
\item Preferential attachment: When choosing the nodes to which the new node connects, we assume that the
probability $\Pi $ that a new node will be connected to node $i$
depends on the degree $k_i$ of node $i$, such that
\begin{equation}
\Pi(k_i)=\frac{k_i}{\sum_j k_j}.
\end{equation}
\end{enumerate}
Scale-free network generated in such way has an exponent
$\gamma=3$. It is well known that the BA model has small
clustering coefficient $C$, which decreases with the network size,
following approximately a power law $C\sim N^{-0.75}$.

Figure 3 is the average time for various search strategies on the BA models of different size $N$. It is notable
that all the linear lines are almost parallel. In particular, the lines for NB, NTL, and NQL collapse into one,
implying clustering effects are quite limit. On the other hand, the improvement of search efficiency for the SA
random walk is related to the small-world feature of the SF network. In this figure we also plot the search time
for a preferentially self-avoid random walk (dashed line). For the PSA walk, we design that the hopping
probability of the walker to a near neighbor is proportional to the degree of this neighbor. Hence a star node is
in favored position. However, one can see that this walk strategy does not work as good as the primitive
self-avoid random walk. At first sight, this result is somehow contrary to one's intuition. It is also at variance
with a conclusion obtained in Ref.\cite{Adamic}, which claims that the optimal search strategy is the PSA walk.
For this point, we argue that although from star nodes a walker may conveniently get to more nodes, these star
nodes are more frequently revisited from other nodes. Hence PSA walk may double-count the advantage of star nodes.
In other words, the star nodes are too frequently visited. It becomes more difficult to reach a target on nodes of
small degree. The same conclusion applies to other networks. It should be mentioned that Ref.\cite{Adamic} assumed
that the walker knows not only the degree of nearest neighbors, but also the information of neighbors' neighbors.
This is a rather strict condition for a random walker. I am not sure if the variance between us originates from
this additional assumption.

To reveal the characteristics of the large connectivity
heterogeneity of scale-free networks, we study the mean
first-passage time (MFPT) $\langle T_j\rangle $ of a given nodes
$j$ with $K_j$ near neighbors. Suppose the walker starts at node
$i$ at time $t=0$, then the master equation for the probability
$P_{ij}$ to find the walker at node $j$ at time $t$ is

\begin{equation}
P_{ij}(t+1)=\sum_k \frac{A_{ij}}{K_k}P_{ik}(t).
\end{equation}
According to Ref.\cite{Noh}, one can obtain that
\begin{equation}
\langle T_j\rangle\propto 1/K_j.
\end{equation}

Namely, nodes with larger degrees is more easily to be found out than nodes with smaller degrees. At a mean field
level, They are preferred in receiving information from the whole network. It is notable that MFPT is independent
of the explicit topological structure of the underlying networks. In Fig.4 we depict numerical simulations of the
MFPT according to the degree distribution for a system of size $N=1000$. After average over the search times for
given degree $k$'s (the lower panel), the fitted line shows that it is indeed inversely proportional to the
corresponding node degree.

We note that some nodes with small number of neighbors also have small MFPT (Fig.4, upper panel). These nodes play
important roles in dynamical processes of information propagation. It is called the {\it random-walk
betweenness}\cite{Noh}, which is not necessary the same as the shortest-path betweenness extensively discussed in
literature.

\subsection{Clustered scale-free models}
One major deficit of the BA model is the lack of high clustering coefficient, which is present in most practical
networks. In particular, the clustering coefficient usually has a degree-dependent power-law form $C(k)\propto
k^{-\gamma}$, with $\gamma\sim 1$ for most cases\cite{Jeong,Ravasz,Maslov,Newman,BJN}. Lots of models have been
proposed to account for this hierarchical feature\cite{Bianconi,Krapivski,DGM,Ravasz,Holme,Klemm}. As an example,
we will consider the growing model (the deactivation model) introduced by Klemm and Egu\'{i}luz in which nodes are
progressively deactivated with a probability inversely proportional to their connectivity\cite{Klemm}. Analytical
arguments and numerical simulations have lead to the claim that, under general conditions, the deactivated model
allowing a core of $m$ active nodes, generating a network with average degree $\langle k\rangle =2m$ and degree
probability distribution $P(k)=2m^2k^{-3}$. The scale-free properties are associated to a high clustering
coefficient.

The model starts from a completely connected graph of $m=2$ active
nodes and proceeds by adding new nodes one by one. Each time a
node is added, (1) it is connected to all active nodes in the
network; (2) one of the active nodes is selected and set inactive
with probability
\begin{equation}
p_d(k_i)=\frac{[\sum_{j\in A}(a+k_j)^{-1}]^{-1}}{a+k_i};
\label{deactive}
\end{equation}
and (3) the new node is set active. The sum in Eq.(\ref{deactive})
runs over the set of active nodes $A$. $a=2$ is a model parameter.

Figure 5 is a plot as that in Fig.3. Under this circumstance, one finds that NTL walk and NQL walk all improve the
search efficiency, in contrast to the results in the BA model. This is resulted from the large clustering
phenomena of the deactivation model. As usual, the further reduction of search time for the self-avoid random walk
reflects the small world property of this model.

\section{summary}
We have played random walks on complex architectures from random graph, small-world to scale-free networks. The
walker may take different walking strategies to promote search efficiency. It is found that the self-avoid random
walk is the most efficient search strategy if the walker is not aware of the global structure of the underlying
network. NTL and NQL walks can be adopted to probe the clustering phenomena of the networks by comparing with the
results of RW and NB walk, while SA walk is used as a tool to probe the small-world property by checking if it can
further reduce the average search time. The preferentially SA walk does not help to improve the search process
further. The possible reason is analyzed. We see that dynamical processes on networks are greatly dependent on the
topological features of the networks. In return, it is useful to explore the network topology by employing
appropriate walk strategies.

This work is supported by a grant from Beijing Normal University.

\centerline {Figure Captions}

Figure 1 Log-Log plot of average search time $t$ versus link probability $p$ of a random graph. The total number
of nodes $N=1000$. The slope of the line is 1. All the walk strategies behavior similarly.

Figure 2(Color online) Dependence of average search times on the short-cut probability $p$ on the WS model for RW,
NB, NTL, NQL, and SA walks, respectively. The total number of nodes $N=300$.

Figure 3(Color online) Average search times for the BA model. All the straight lines are parallel to each other.
In particular, the lines for NB, NTL, and NQL collapse into one. The search time of PSA walk (dashed line) is
slightly higher than that of the SA walk.

Figure 4(Color online) Upper panel: MFPT for the BA model for targets with large connectivity heterogeneity. Lower
panel: average over the search times of given $k$'s. A fitted linear relation is obtained with a slope of $-1$.

Figure 5(Color online) Average search times for the deactivation model with high clustering coefficient. The
power-law relations inhibit in all search processes for RW, NB, NTL, NQL, and SA walks, respectively.

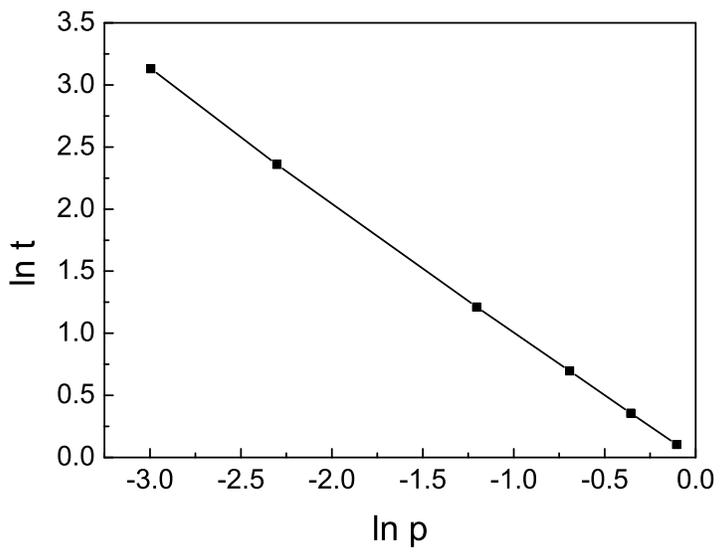
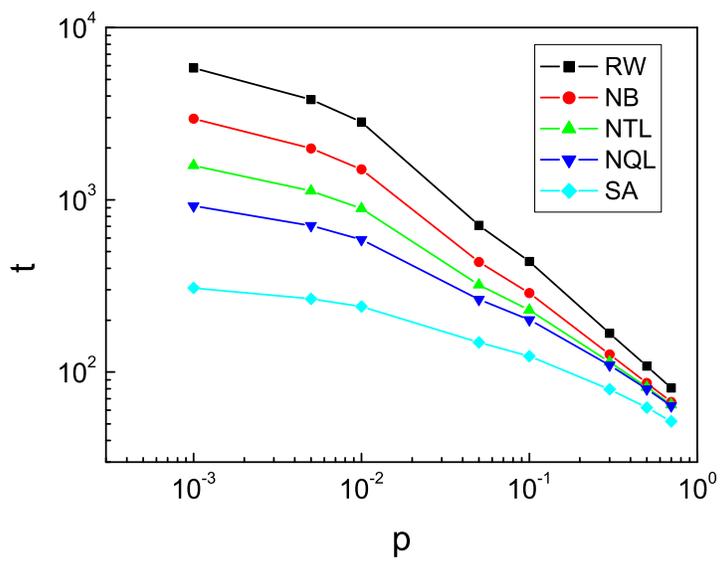
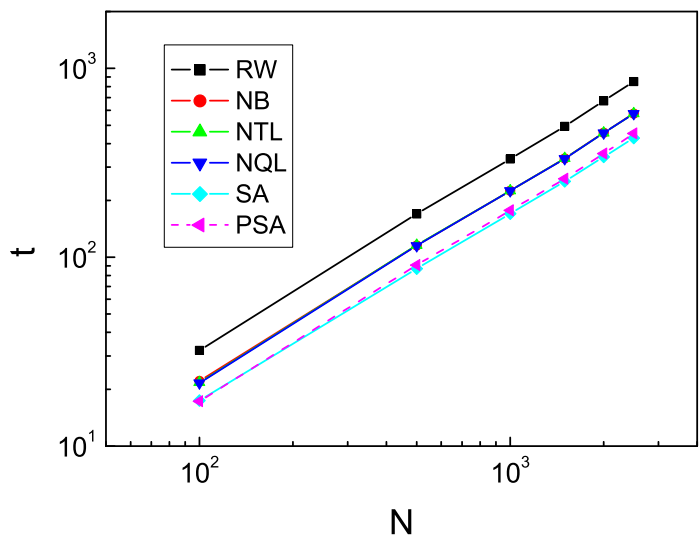
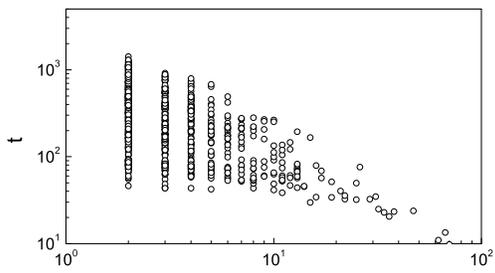
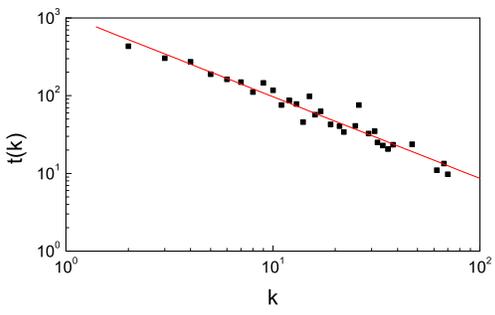
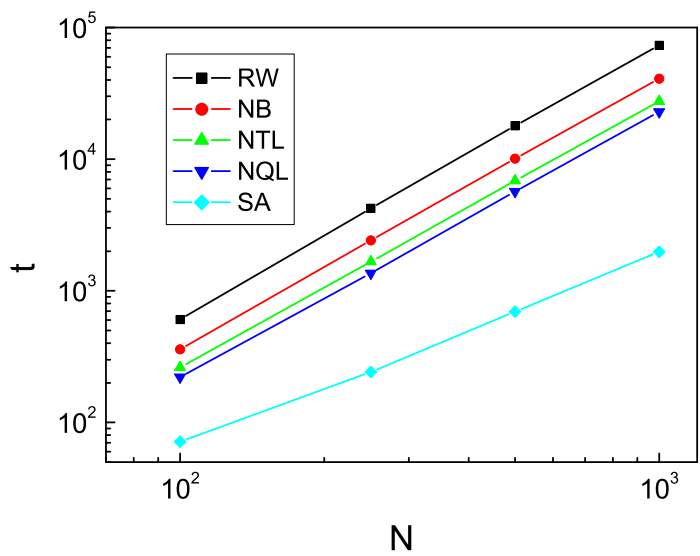
\end{document}